\documentclass[english,aps,manuscript]{revtex4-1}
\usepackage[T1]{fontenc}
\usepackage[latin9]{inputenc}
\usepackage{float}
\usepackage{graphicx}

\makeatletter

\providecommand{\tabularnewline}{\\}

 \@ifundefined{textcolor}{}
 {%
   \definecolor{BLACK}{gray}{0}
   \definecolor{WHITE}{gray}{1}
   \definecolor{RED}{rgb}{1,0,0}
   \definecolor{GREEN}{rgb}{0,1,0}
   \definecolor{BLUE}{rgb}{0,0,1}
   \definecolor{CYAN}{cmyk}{1,0,0,0}
   \definecolor{MAGENTA}{cmyk}{0,1,0,0}
   \definecolor{YELLOW}{cmyk}{0,0,1,0}
 }

\usepackage{babel}

\makeatother

\usepackage{babel}
\begin{document}

\title{Majorana Neutrino and $W_{R}$ at TeV scale $ep$ Colliders}

\author{U. Kaya}

\address{Ankara University, Faculty of Science, Department of Physics, 06100
Tando\u{g}an, Ankara, TURKEY}

\email{umit.kaya@cern.ch}

\author{M. Sahin}

\address{Usak University, Science and Letters Faculty Department of Physics,
64100, Usak, TURKEY}

\email{mehmet.sahin@usak.edu.tr}

\author{S. Sultansoy}

\address{TOBB Economy and Technology University, 06520 Sogutozu, Ankara, TURKEY}

\address{National Academy of Sciences, Institute of Physics, Baku, Azerbaijan}

\email{s.sultansoy@etu.edu.tr}

\begin{abstract}
Production of heavy Majorana neutrino $N_{e}$ predicted by left-right
symmetric extension of the Standard Model at future {\normalsize TeV}
scale $ep$ colliders have been considered. In order to estimate potential
of $ep$ colliders for $N_{e}$ search we consider back-groundless
process $e^{-}p\rightarrow e^{+}X$ which is consequence of Majorana
nature of $N_{e}$. It is shown that {\normalsize linac-LHC} and {\normalsize linac-FCC}
based $ep$ colliders will cover much wider regions of $N_{e}$ and
$W_{R}$ masses than corresponding linear electron-positron colliders. 
\end{abstract}
\maketitle

\section*{1. Introduct\i{}on}

The discovery of the Higgs boson \cite{ATLAS Collaboration,CMS Collaboration}
has completed basics of the Standard Model (SM). However, this is
not the end of the story. First of all, SM could not answer many fundamental
questions;for instance, left-handed structure of charged weak currents
is put by hand, mass and mixing pattern of the SM fermions are not
clarified etc. For these reasons a number of Beyond the SM (BSM) models
are developed. Fortunately, future running of the LHC with upgraded
parameters will shed light on answers for some of these questions.
Secondly, Higgs field provides only $2\%$ of the mass of the visible
universe. In order to understand the origin of remaining $98\%$ we
should clarify basics of the QCD, especially small-$x$ (Bjorken)
region. This is why the QCD Explorer option of the LHC-based $ep$
colliders is mandatory \cite{Cetin}.

TeV scale colliders can be classified using colliding particles or
collider types. While the first classification includes hadron-hadron,
lepton-hadron and lepton-lepton collisions the other includes collider
types with ring-ring, linac-ring and linac-linac options. Concerning
the energy frontiers linac-ring type colliders provide the sole realistic
way to handle (multi) TeV scale in lepton-hadron collisions at constituent
level \cite{Sultansoy} (see, also, review \cite{Akay}). Certainly,
hadron colliders (LHC and FCC) should be considered as the main discovery
tools for new particles. In this respect, priority of lepton-lepton
or lepton-hadron colliders will be determined by the LHC or FCC results.
If the observed BSM physics will be connected to the first family
fermions, $ep$ colliders will have priority due to larger center-of-mass
energy comparing to electron-positron colliders.

Left-handed structure of charged weak currents has most elegant explanation
in $SU_{L}(2)\times SU_{R}(2)\times U(1)_{B-L}$ models \cite{Pati},
where the mass of $W_{R}$ boson is much higher than the mass of $W_{L}$
boson. In addition, these models predict existence of the second heavy
$Z$ boson, as well as, three heavy Majorana neutrinos ($N_{e}$,
$N_{\mu}$, $N_{\tau}$) and a number of additional Higgs bosons,
including double-charged ones. Therefore, there are a lot of interesting
phenomena, that can be investigated at TeV energy colliders. For example,
production of $W_{R}$ at hadron colliders, followed by $W_{R}\rightarrow N_{l}\, l$
decays, leads to a very clear signature, namely two same-sign leptons
in association with two jets \cite{Keung}. Recently, the process
$pp\rightarrow W_{R}+X$ followed by $W_{R}\rightarrow l\, l\, j\, j$
decay has been investigated by CMS collaboration \cite{CMS Collaboration2},
where $2.8\sigma$ excess at $m(eejj)=2.1$ TeV is observed. However,
only one of the $14$ reconstructed events contains same-sign electrons,
while it is expected half of the events to be same-sign in the case
of Majorana $N_{e}$.

In this paper, we analyze production of $N_{e}$ at future $ep$ colliders
via the back-groundless process $e^{-}p\rightarrow e^{+}X$. In Section
2 tentative parameters of possible LHC and FCC based $ep$ colliders
are presented. A brief description of the $SU_{L}(2)\times SU_{R}(2)\times U(1)_{B-L}$
model is given in Section 3. In Section 4, we show production cross-sections
depending on $N_{e}$ mass for several values of $W_{R}$ mass and
different center-of-mass energies, as well as $3\sigma$ and $5\sigma$
plots in $N_{e}$-$W_{R}$ mass plane. Rough comparison of $pp$,
$ep$ and $e^{+}e^{-}$ colliders potentials for $N_{e}$ investigation
is performed in Section 5. Finally, in Section 6 we give some conclusions
and recommendations.

\section*{2. Future \textup{\normalsize TeV}{\normalsize {} scale $ep$ colliders}}

Electron-proton scattering experiments have played a very important
role in the development of our knowledge of the structure of matter.
For example, quark-parton structure was discovered in deep-inelastic
scattering. As mentioned above, QCD Explorer option ($E_{e}=60$ GeV)
of the LHC-based $ep$ colliders (LHeC) is mandatory for clarifying
dynamics of strong interactions at extremely small-$x$ and, at the
same time, sufficiently high $Q^{2}$ region. In addition, EIC (e-RHIC)
will give opportunity to investigate another critical region, namely,
$x\approx1$.

Concerning BSM, new physics higher electron and/or proton beam energies
are needed. In Table 1 we present tentative parameters for possible
LHC and FCC based $ep$ collider options. OPL, ERL and OPERL denotes
one pulse linac, energy recovery linac and one pulse energy recovery
linac, respectively. ERL60 is continuous wave superconducting recirculating
energy-recovery linac, which has been chosen as the main line for
LHeC (see section 7.1.2 in LHeC CDR \cite{Fernandez}). OPL60 and
OPL140 options for LHeC are considered in Section 7.1.4 of the LHeC
CDR, and OPERL150 option is considered in Section 7.1.5, where this
option is named as higher-energy LHeC ERL option. Luminosity for OPL500
and OPERL500 options for LHeC are estimated by assuming that electron
beam powers are the same as in the OPL140 and OPERL150 cases, respectively
(in this case $L$ is inverse proportional to $E_{e}$). Parameters
for the first three options of FCC-based $ep$ colliders are taken
from \cite{Zimmermann}. Luminosity values for the last two options
are taken same as OPL140 and OPERL150 LHeC options for following reason:
the loss due to higher electron energy $E_{e}$ is compensated by
smaller proton beam size due to higher $E_{p}$.

\begin{table}[H]
\caption{Parameters of possible LHC and FCC based $ep$ colliders.}

\centering{}%
\begin{tabular}{|c|c|c|c|c|c|}
\hline 
 & $E_{e}$, GeV  & $E_{p}$, TeV  & $\sqrt{s}$, TeV  & $L$, $10^{33}\, cm^{-2}\, s^{-1}$  & $L_{int}$, $fb^{-1}(10^{7}\, s)$\tabularnewline
\hline 
\hline 
ERL60-LHC  & $60$  & $7$  & $1.3$  & $1$$([9])$  & $10$\tabularnewline
\hline 
OPL60-LHC  & $60$  & $7$  & $1.3$  & $0.09$$([9])$  & $0.9$\tabularnewline
\hline 
OPL140-LHC  & $140$  & $7$  & $2.0$  & $0.04$$([9])$  & $0.4$\tabularnewline
\hline 
OPL500-LHC  & $500$  & $7$  & $3.7$  & $0.01$  & $0.1$\tabularnewline
\hline 
OPERL150-LHC  & $150$  & $7$  & $2.0$  & $100$$([9])$  & $1000$\tabularnewline
\hline 
OPERL500-LHC  & $500$  & $7$  & $3.7$  & $30$  & $300$\tabularnewline
\hline 
ERL60-FCC  & $60$  & $50$  & $3.5$  & $10$$([10])$  & $100$\tabularnewline
\hline 
FCC-e80  & $80$  & $50$  & $4.0$  & $23$$([10])$  & $230$\tabularnewline
\hline 
FCC-e120  & $120$  & $50$  & $4.9$  & $12$$([10])$  & $120$\tabularnewline
\hline 
OPL1000-FCC  & $1000$  & $50$  & $14.1$  & $0.04$  & $0.4$\tabularnewline
\hline 
OPERL1000-FCC  & $1000$  & $50$  & $14.1$  & $100$  & $1000$\tabularnewline
\hline 
\end{tabular}
\end{table}

\section*{3. Model Description}

The minimal left-right symmetric model is based on $SU(3)_{C}\times SU(2)_{L}\times SU(2)_{R}\times U(1)_{B-L}$
gauge symmetry. Fermions are grouped in $L-R$ symmetric form:

\begin{equation}
q_{L}=\left(\begin{array}{c}
u\\
d
\end{array}\right)_{L}:(3,2,1,1/3);\; q_{R}=\left(\begin{array}{c}
u\\
d
\end{array}\right)_{R}:(3,1,2,1/3);
\end{equation}

$\,$

\begin{equation}
L_{L}=\left(\begin{array}{c}
\nu\\
l
\end{array}\right)_{L}:(1,2,1,-1);\; L_{R}=\left(\begin{array}{c}
N\\
l
\end{array}\right)_{R}:(1,1,2,-1).
\end{equation}

$\,$

In this model electric charge is given by

\begin{equation}
Q_{em}=I_{3L}+I_{3R}+\frac{B-L}{2}
\end{equation}

$\,$

The Higgs sector contains one bidoublet and two triplets:

\begin{equation}
\Phi=\left(\begin{array}{cc}
\phi_{1}^{0} & \phi_{1}^{+}\\
\phi_{2}^{-} & \phi_{2}^{0}
\end{array}\right):(1,2,2,0);\;\Delta_{L}=\left(\begin{array}{cc}
\delta_{L}^{+}/\sqrt{2} & \delta_{L}^{++}\\
\delta_{L}^{0} & -\delta_{L}^{+}/\sqrt{2}
\end{array}\right):(1,3,1,2);
\end{equation}

$\,$

\begin{equation}
\Delta_{R}=\left(\begin{array}{cc}
\delta_{R}^{+}/\sqrt{2} & \delta_{R}^{++}\\
\delta_{R}^{0} & -\delta_{R}^{+}/\sqrt{2}
\end{array}\right):(1,1,3,2).
\end{equation}

$\,$

Therefore, there are two additional intermediate vector bosons ($W_{R}$
and $Z_{R}$) as well as a number of additional scalar bosons (2 neutral,
2 charged and 2 double charged). The most stringent limit on $W_{R}$
mass from low energy data, namely, $m{}_{W_{R}}>2.5$ TeV, comes from
$K_{L}-K_{S}$ mixing \cite{Zhang}. Search for $W_{R}\rightarrow lN_{l}$
performed by ATLAS and CMS using $7$ TeV data lead to the same limit
for $m{}_{N_{l}}$ around $1$ TeV \cite{ATLAS Collaboration2,CMS Collaboration3}.
Recent CMS analysis of this channel increases the limit up to $3$
TeV \cite{CMS Collaboration2}.

\section*{4. Production of Majorana neutrino at \textup{\normalsize TeV}{\normalsize {}
energy $ep$ colliders}}

For numerical calculations we include interactions of $W_{R}$ boson
with fermions, namely,

\begin{equation}
\mathcal{L}=\frac{-g_{R}}{2\sqrt{2}}\left[\bar{e}\gamma^{\mu}\left(1+\gamma^{5}\right)\nu+\bar{d}\gamma^{\mu}\left(1+\gamma^{5}\right)u\right]W_{R\mu}^{-}+h.c.
\end{equation}

\noindent into the CalcHEP software \cite{Belyaev}. At this stage
we ignore possible mixing in fermions and vector boson sectors. In
Fig. 1 we present production cross-sections as a function of $m_{N}$
for $\sqrt{s}=1.3$ TeV and different values of $m{}_{W_{R}}$. Similar
distributions for $\sqrt{s}=2.0,\,4.0$ and $14.1$ TeV are presented
in Figs.$2$, $3$ and $4$, respectively. Keeping in mind, integral
luminosity values given in the last column of the Table $1$, one
can see that ERL60, OPL60 and OPL140 options of the LHC based $ep$
colliders could not provide essential contributions to the $N_{e}$
investigation.

\begin{figure}
\begin{centering}
\includegraphics[scale=0.5]{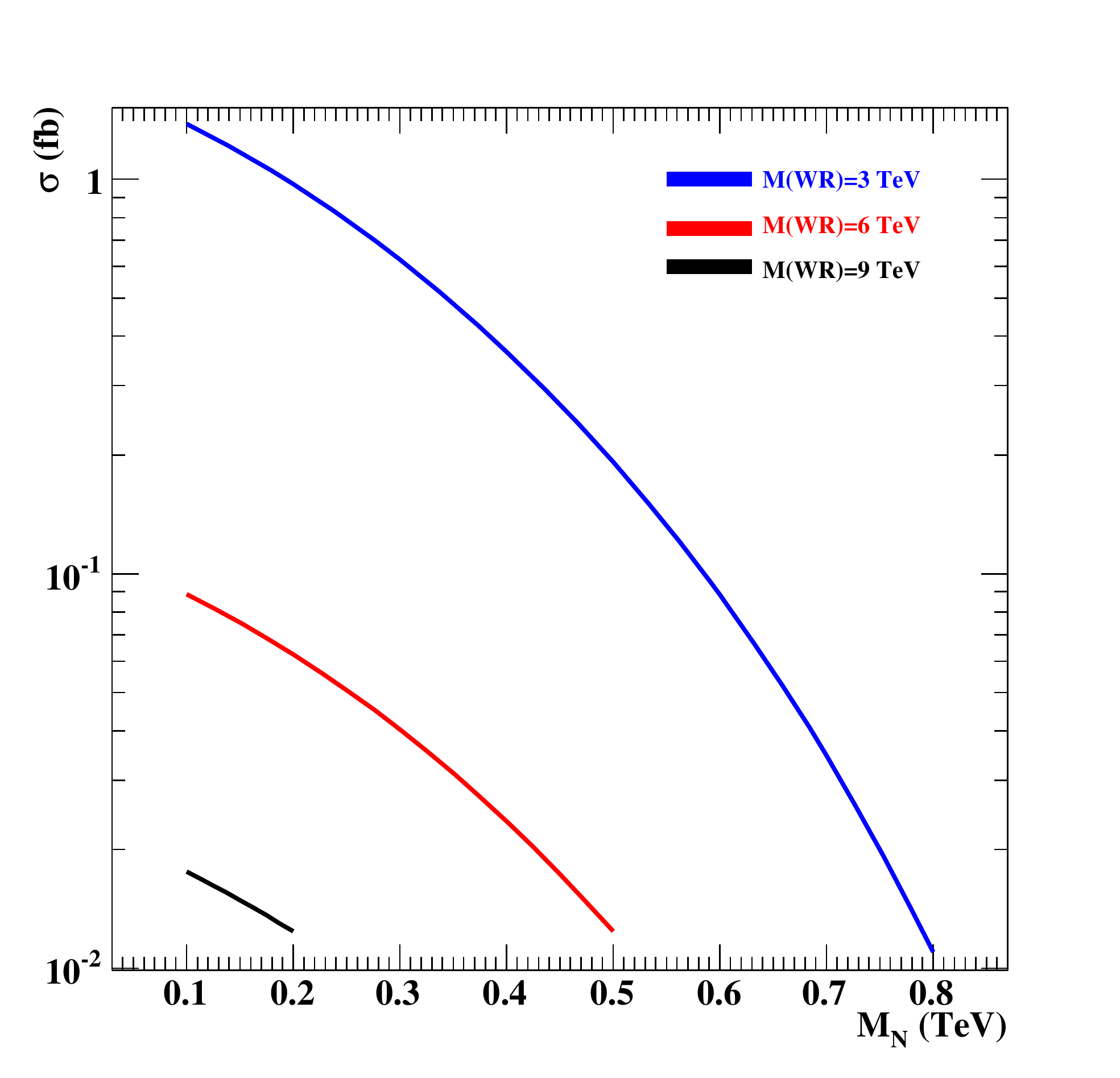} 
\par\end{centering}

\raggedright{}\caption{$N_{e}$ production cross-sections for different $m{}_{W_{R}}$ at
$\sqrt{s}=1.3\, TeV$}
\end{figure}

\begin{figure}
\begin{centering}
\includegraphics[scale=0.5]{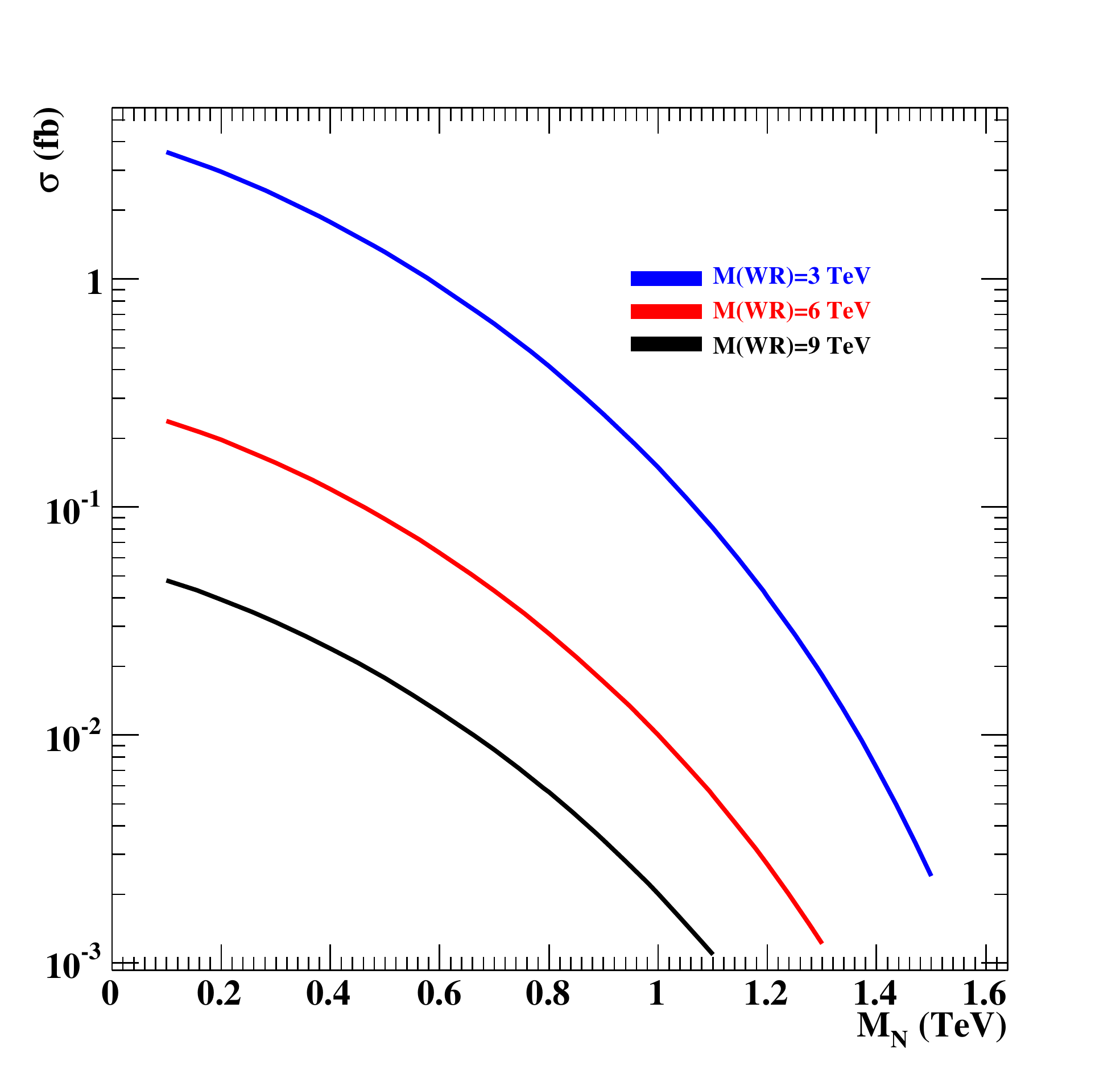} 
\par\end{centering}

\caption{$N_{e}$ production cross-sections for different $m{}_{W_{R}}$ at
$\sqrt{s}=2.0\, TeV$}
\end{figure}

\begin{figure}
\begin{centering}
\includegraphics[scale=0.5]{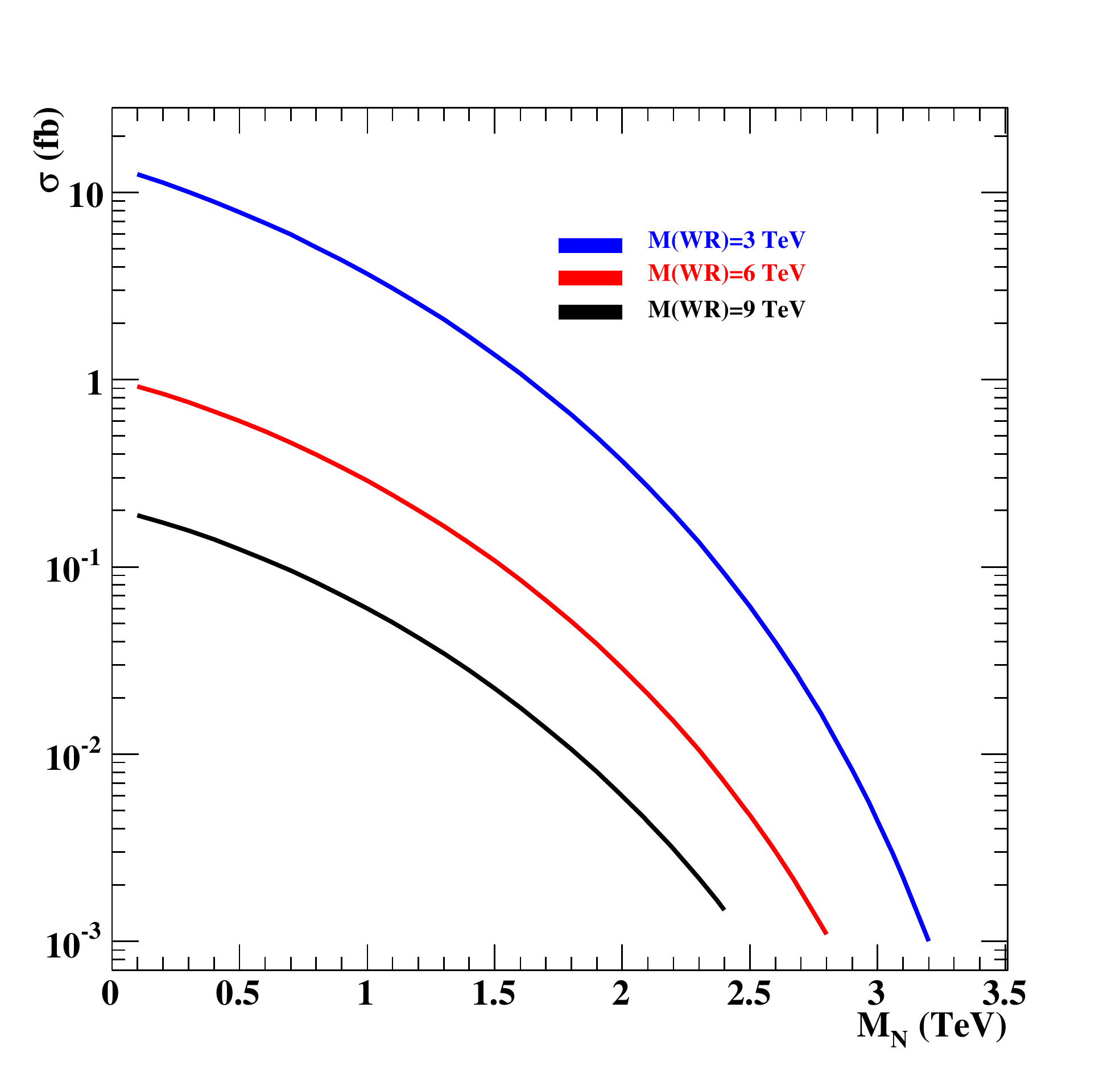} 
\par\end{centering}

\caption{$N_{e}$ production cross-sections for different $m{}_{W_{R}}$ at
$\sqrt{s}=4.0\, TeV$}
\end{figure}

\begin{figure}
\begin{centering}
\includegraphics[scale=0.5]{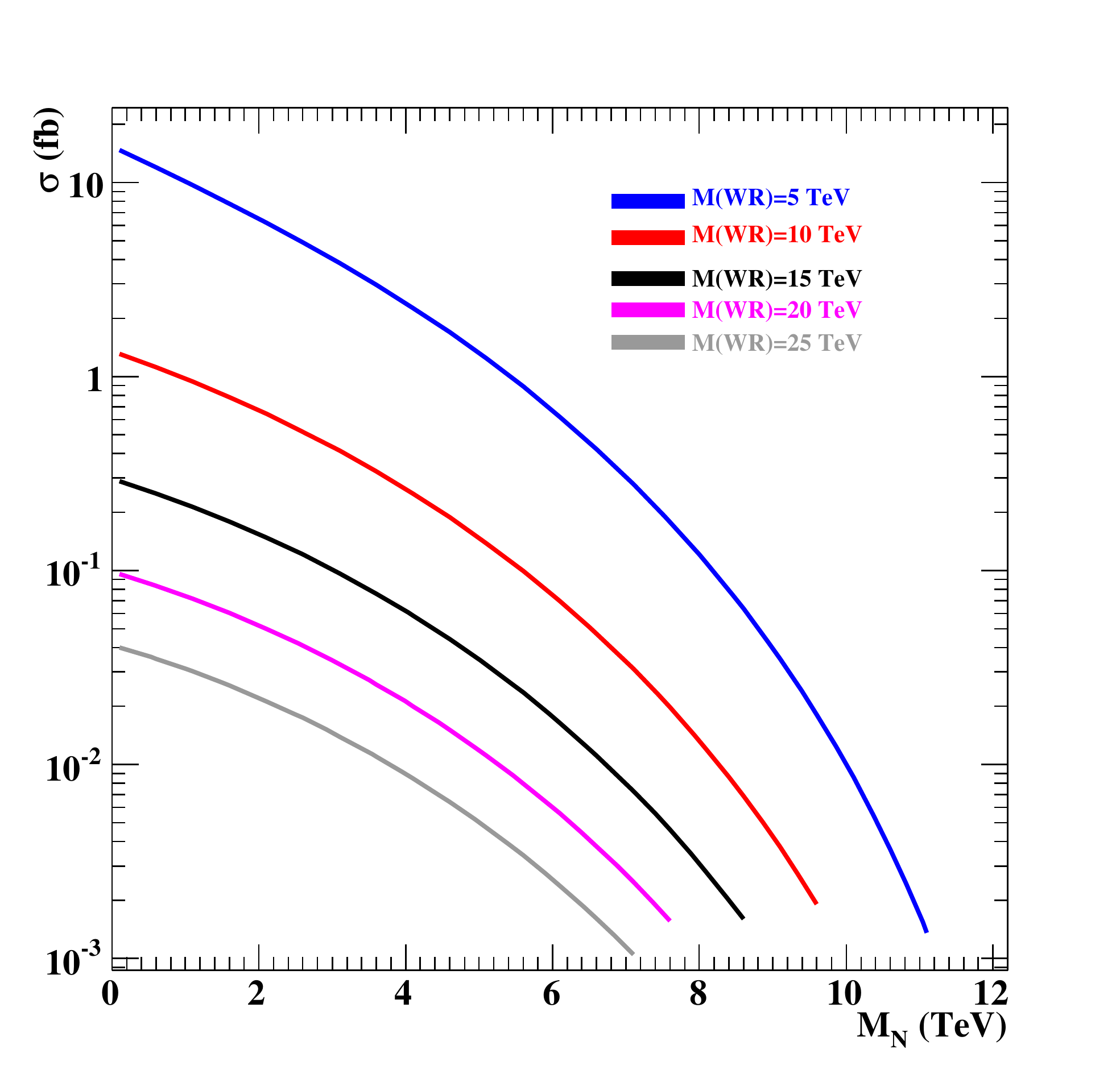} 
\par\end{centering}

\caption{$N_{e}$ production cross-sections for different $m{}_{W_{R}}$ at
$\sqrt{s}=14.1\, TeV$}
\end{figure}

In order to estimate the potential of $ep$ colliders for $N_{e}$
search, we consider back-groundless process $e^{-}p\rightarrow e^{+}X$
which is consequence of Majorana nature of $N_{e}$. Corresponding
Feynman diagram is shown in Fig. $5$. In this case cross-sections
given in Figs. $1-4$ should be multiplied by factor $1/2$. In Figs.
$6-9$, we present $3\sigma$ (observation) and $5\sigma$ (discovery)
plots in $m_{N_{e}}-m{}_{W_{R}}$ plane. For $3\sigma$ and $5\sigma$
limits we use $9$ and $25$ events, respectively. As one can see
from Fig. $6$, LHC based $ep$ colliders could probe larger $m{}_{W_{R}}$
mass values than LHC if $N_{e}$ has relatively light mass.

\begin{figure}
\begin{centering}
\includegraphics[scale=0.9]{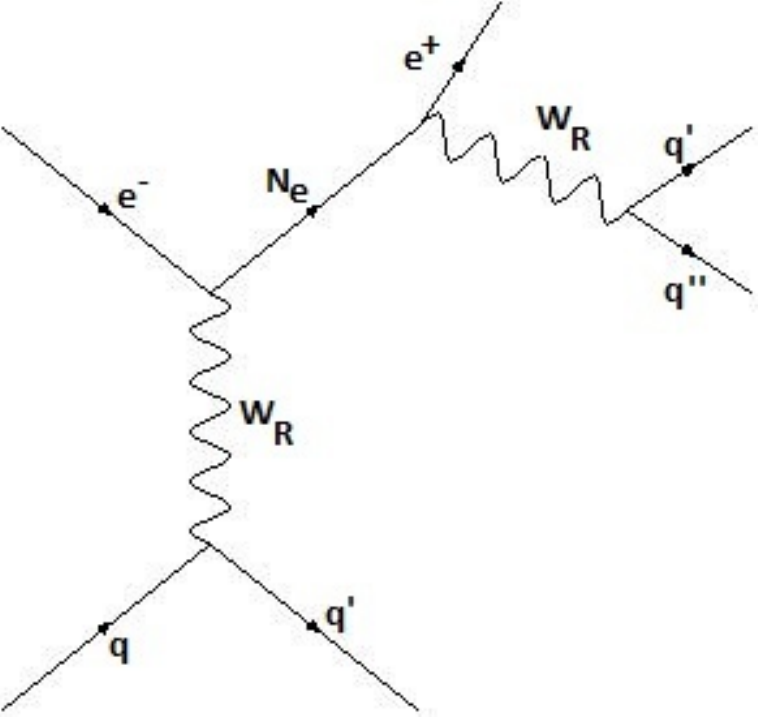} 
\par\end{centering}

\caption{Feynmann diagram for $e^{-}p\rightarrow e^{+}+X$}
\end{figure}

\begin{figure}
\begin{centering}
\includegraphics[scale=0.5]{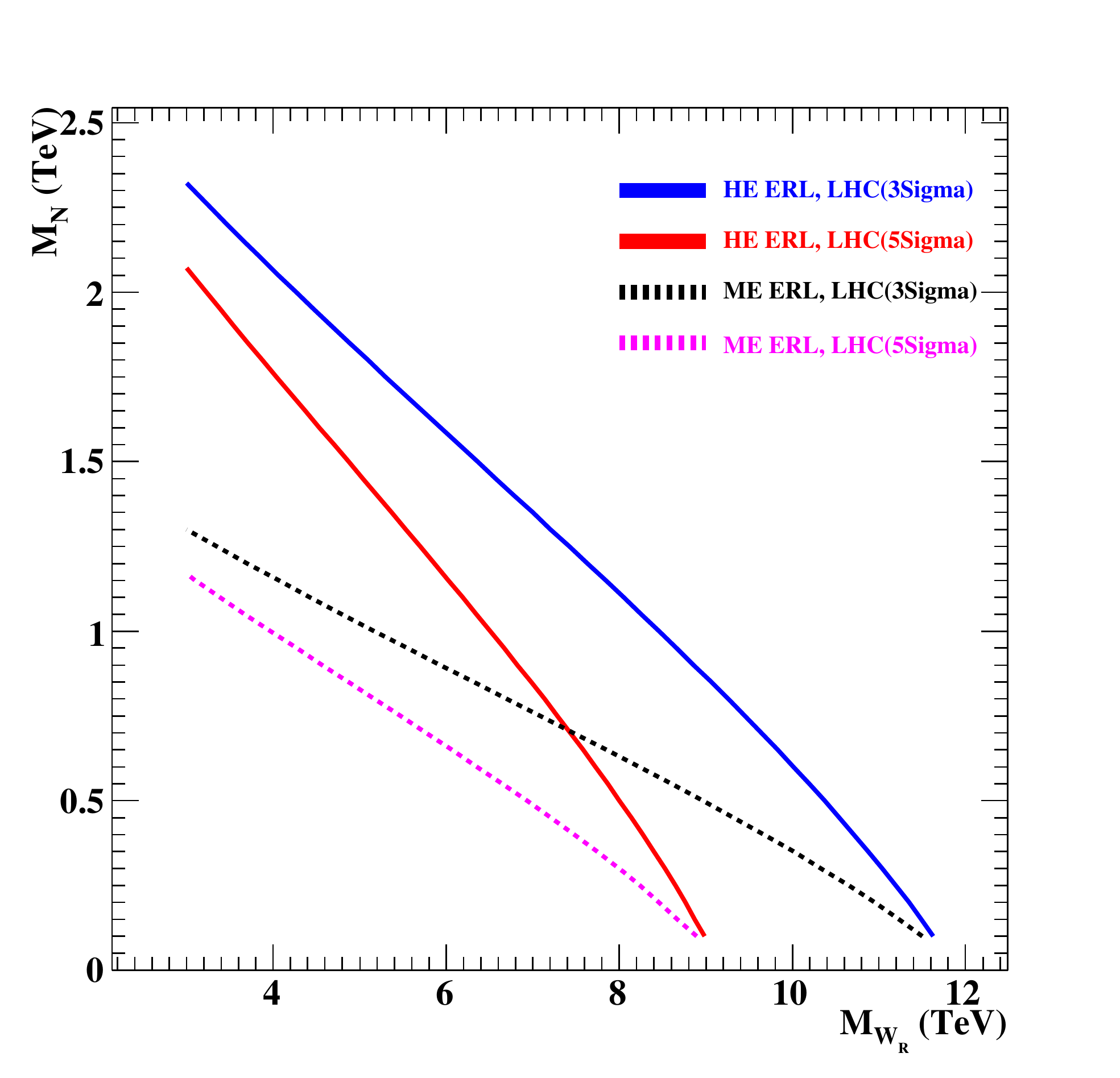} 
\par\end{centering}

\caption{Observation ($3\sigma$) and discovery ($5\sigma$) limits in the
$m_{N_{e}}-m{}_{W_{R}}$ plane for OPERL$150$ and OPERL$500$ options
of LHeC.}
\end{figure}

\begin{center}
\begin{figure}[H]
\begin{centering}
\includegraphics[scale=0.5]{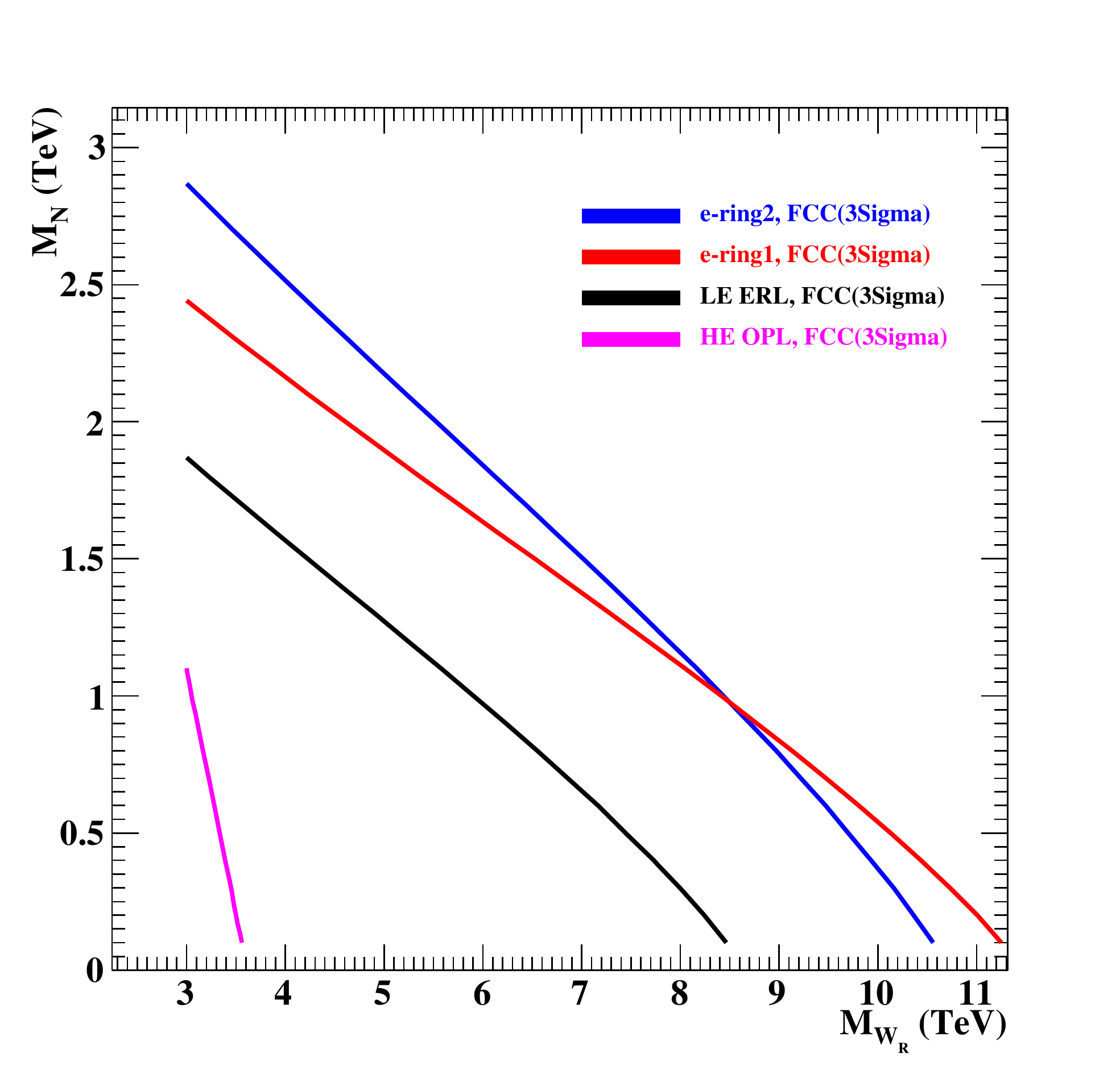} 
\par\end{centering}

\caption{Observation ($3\sigma$) limit in the $m_{N_{e}}-m{}_{W_{R}}$ plane
for FCC-e80, FCC-e120, ERL60-FCC and OPL1000-FCC. }
\end{figure}

\par\end{center}

\begin{center}
\begin{figure}[H]
\begin{centering}
\includegraphics[scale=0.5]{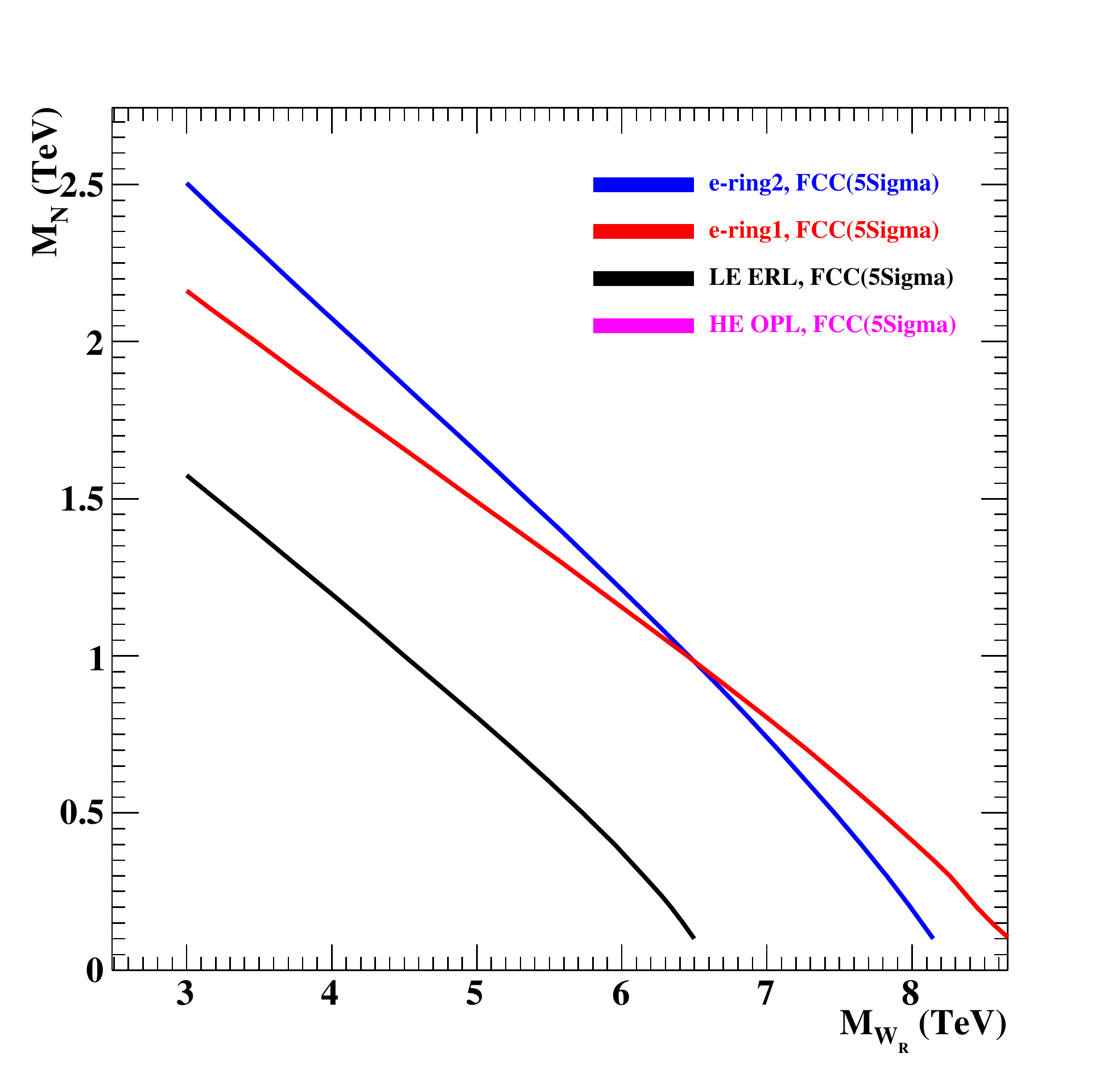} 
\par\end{centering}

\caption{Discovery ($5\sigma$) limit in the $m_{N_{e}}-m{}_{W_{R}}$ plane
for FCC-e80, FCC-e120, ERL60-FCC and OPL1000-FCC.}
\end{figure}

\par\end{center}

\begin{center}
\begin{figure}[H]
\begin{centering}
\includegraphics[scale=0.5]{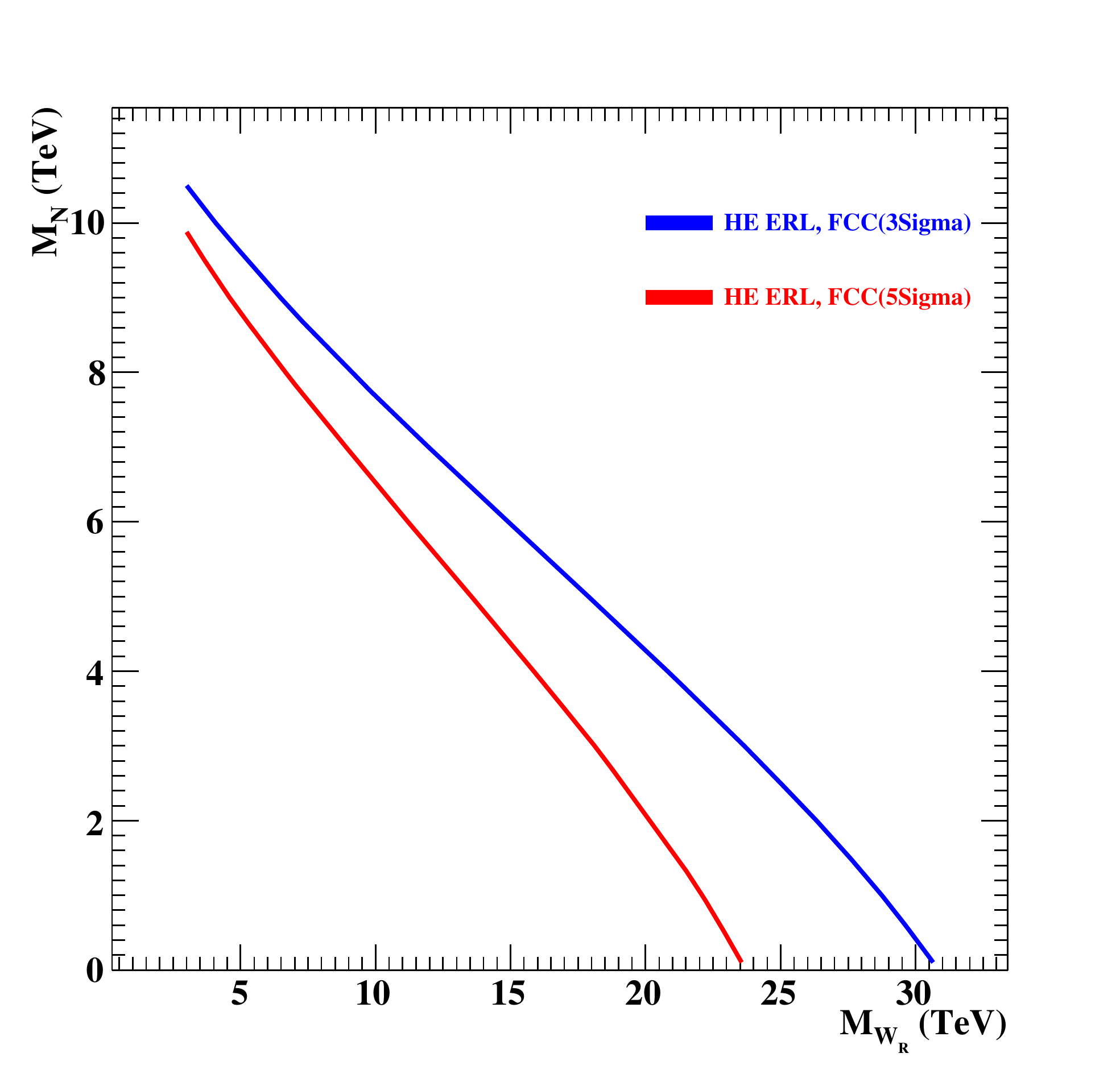} 
\par\end{centering}

\caption{Observation ($3\sigma$) and discovery ($5\sigma$) limits in the
$m_{N_{e}}-m{}_{W_{R}}$ plane for OPERL1000-FCC.}
\end{figure}

\par\end{center}

\section*{5. Rough comparison of $pp$, $ep$ and $e^{+}e^{-}$ colliders}

As it is mentioned in the introduction, TeV scale $ep$ colliders
are crucial for understanding of the origin of $98\%$ of the mass
of the visible universe. Concerning BSM physics, detailed comparison
of physics search potential of (multi)TeV scale hadron, lepton and
lepton-hadron colliders should be performed for each phenomena (such
as new particles and new interactions). Rough estimations \cite{Sultansoy2}
show that $ep$ colliders are advantageous comparing to lepton colliders
for a lot phenomena. Moreover, analysis of excited electron production
performed in the article by O.Cakir et al.\cite{Cakir} \textquotedbl{}show
that LC$\oplus$LHC is more promising than the LHC and much more promising
than the LC for the processes considered\textquotedbl{}. In this analysis
following collider parameters were used: $\sqrt{s}=14$ TeV and $L=10^{34}cm^{-2}s^{-1}$
for LHC, $\sqrt{s}=0.5$ TeV and $L=10^{34}cm^{-2}s^{-1}$ for LC,
$\sqrt{s}=3.74$ TeV and $L=10^{31}cm^{-2}s^{-1}$ for LC$\oplus$LHC.

In principle, for effective planning of the post-LHC energy frontier
colliders comparison of LHC, LHeC and ILC (for 2020s), as well as
FCC ($pp$ and $ep$ options), linac-FCC ($ep$ and $\gamma p$ options),
CLIC and muon collider (for $2030$s), search potentials for different
BSM phenomena should be performed.

Considering $N_{e}$ production, lepton collider capacity is kinematically
limited by $m_{N_{e}}<0.5$ TeV at ILC \cite{ILC} and $m_{N_{e}}<1.5$
TeV at CLIC \cite{CLIC}. As one can see from Figs. $6-9$, linac-LHC
and especially linac-FCC based $ep$ colliders will cover much higher
$m_{N_{e}}$ values. Below we consider specific examples for LHC and
FCC.

1) LHC example: Let us assume that $N_{e}$ and $W_{R}$ masses are
$0.5$ TeV and $8$ TeV, respectively. In this case ERL500-LHC will
give opportunity to discover $N_{e}$ at $5\,\sigma$ level, whereas
LHC will not reach even $2\,\sigma$ level.

2) FCC example: Let us assume that $N_{e}$ and $W_{R}$ masses are
$2$ TeV and $27$ TeV, respectively. In this case, OPERL1000-FCC
will give opportunity to discover Ne at $3\,\sigma$ level, whereas
FCC will not reach even $2\,\sigma$ level.

\section*{6. Conclusion}

In this paper we analyzed production of heavy Majorana neutrino via
$W_{R}$ exchange at future ep colliders. The analysis results show
that:

$1)$ linac-LHC and linac-FCC based ep colliders will cover much wider
regions of $N_{e}$ and $W_{R}$ masses than ILC and CLIC,

$2)$ ep colliders seems to be advantageous comparing to corresponding
pp colliders for some regions of $N_{e}$ and $W_{R}$ masses. 
\begin{acknowledgments}
Authors are grateful to Y. O. Gunaydin for useful discussion and valuable
comments.\end{acknowledgments}

\end{document}